\documentclass[10pt]{wlscirep}
\usepackage[utf8]{inputenc}
\usepackage[T1]{fontenc}

\usepackage{float}
\usepackage{amsmath}
\title{Depth-resolved polarisation sensitive optical coherence tomography reveals the complex microanatomical response of cartilage to compression}

\author[1,2,*]{Darven Murali Tharan}
\author[1,2]{Marco Bonesi}
\author[2,3]{Daniel Everett}
\author[1,2]{Matthew Goodwin}
\author[1,2]{Cushla McGoverin}
\author[4]{Sue McGlashan}
\author[3]{Ashvin Thambyah}
\author[1,2]{Frédérique Vanholsbeeck}

\affil[1]{The University of Auckland, Department of Physics, Auckland, New Zealand, 1010}
\affil[2]{The Dodd Walls Centre for Quantum and Photonic Technology}
\affil[3]{The University of Auckland, Department of Chemical and Materials Engineering, Auckland, New Zealand, 1010}
\affil[4]{The University of Auckland, Department of Anatomy and Medical Imaging, Auckland, New Zealand, 1010}

\affil[*]{darven.murali.tharan@auckland.ac.nz}

\begin{abstract}
Conventional methods for analysing cartilage microstructure under mechanical loading are largely destructive. In this work, we evaluate the efficacy of using depth-resolved polarisation sensitive optical coherence tomography (PS-OCT) to study the cartilage morphological response to compression. We show that depth-resolved PS-OCT reveals the microstructure of cartilage under load, and it can do so non-destructively, opening significant possibilities for enhanced clinical assessment of cartilage health by detecting deviance from normal load-bearing behaviour.
\end{abstract}

\begin{document}

\flushbottom
\maketitle
\thispagestyle{empty}

\section{Introduction}

Articular cartilage is a specialised connective tissue that plays a crucial role in maintaining joint function by distributing loads and minimising friction during movement. Cartilage consists largely of water and a dense collagen network that constrains hydrophilic proteoglycan. These proteoglycans attract water, generating immense hydrostatic pressure as required. This pressure-driven fluid movement through the permeable cartilage matrix contributes to the remarkable load-bearing capabilities of articular cartilage~\cite{maroudas1979physicochemical}.

In contrast to its deceptively simple appearance, being a pale, relatively thin tissue covering the bone ends in articulating joints, cartilage has a complex microstructure critical for its important load-bearing function. In the uppermost region (approximately 10\% of tissue thickness) is a layer that consists of a planar, tangentially aligned, collagen fibrillar network. The collagen network is radially aligned in the middle to the deeper cartilage zone. A `transitional’ zone in between the tangential and radial zones sees the collagen gradually `arching’ from the former orientation to the latter. The overall `zonally differentiated’~\cite{brown2020mechanical} fibrillar architecture in full-thickness cartilage had been described much earlier and named after the author as the `Benninghoff arcades’~\cite{benninghoff1925form}. The deep layers of tissue integrate with the underlying hard bone via a gradual compositional and mechanical transition. Relatively soft articular cartilage graduates to a mineralised region of about a few hundred micrometers thick~\cite{hargrave2021micro,oegema1997interaction} before integrating with the hard subchondral plate bone. 

The different zones of cartilage tissue, including the mineralised region, with their own unique structure, have different mechanical properties~\cite{verteramo2004zonal,hargrave2015multi}, allowing cartilage to effectively distribute loads in a manner unlike any known engineered material. For example, \textit{in vitro} experiments have shown how load distribution involves the surface layer acting like a `strain-limiter’~\cite{thambyah2006micro},  taking stresses away from the directly loaded region. In the deeper zones, there is radial compression of the fibrillar network, resulting in `lateral bulging’~\cite{glaser2002functional,hargrave2021micro} of the tissue. Finally, at the cartilage-bone interface, an intense shear arises from the lateral bulge of the cartilage matrix against a rigid bone substrate. All these different responses, taken together, result in a unique `chevron-like’ pattern of the compressed collagen network. The chevron pattern, with its apex denoted as the shear discontinuity (SD), is thus the result of the competing effects of a strain-limiting tangential zone, against the lateral bulging of the mid-to-deep zones, with the relative compliance of the transition zone. 

Studies of cartilage morphology under load have been dominated by a myriad of destructive analyses~\cite{broom1980study,notzli1997deformation,glaser2002functional,thambyah2006micro,nickien2013changes,kaab2003deformation}. Of note, is the seminal work of Broom \textit{et al}.~\cite{broom1980study}, who employed a mechanical compression device alongside differential interference contrast (DIC) microscopy to investigate the dynamic, `live’ response of thin 2D cartilage slices under load. Most work forgoes observation of the `live’ response, instead focusing on studying cartilage that has been fixed in its compressed state. Notzli and Clark~\cite{notzli1997deformation}, for example, utilised cryofreezing to preserve cartilage in its deformed state under load, later examining its ultrastructural alterations using scanning electron microscopy (SEM). More contemporarily, Thambyah  \textit{et al}.~\cite{thambyah2006micro} utilised DIC microscopy to study cartilage on bone under static load revealing a complex fibrillar response to compression described in the 2nd paragraph above. 

Looking back at the past half-century of cartilage mechanics research, two key challenges remain unresolved. The first is the inability to directly observe the dynamic, live response of cartilage under load, a limitation that, to date, only Broom has meaningfully attempted to address. While Broom’s work highlighted the critical insights gained from studying the cartilage live response, the methodology remains fundamentally flawed, requiring cartilage to be placed in an extremely unnatural and non-physiological state.  As Broom stressed~\cite{broom1980study}, soft tissues undergo large strains and profound ultrastructural changes under load, and without simultaneous observation of both strain and morphology, any interpretation of cartilage mechanics is obscured.

The second challenge lies in the reliance on destructive techniques for morphological analysis, limiting their applicability to fundamental research into cartilage behaviour. Histopathology has established a clear correlation between osteoarthritic cartilage and tissue subjected to repetitive excessive mechanical loading, reinforcing the connection between cartilage health and mechanical behaviour~\cite{zimmerman1988mechanical,radin1991mechanical}. It has been further demonstrated that distinct differences in the loaded morphology of cartilage are observed at different stages of degeneration. For example, the SD effectively disappears when the cartilage is compressed, if the cartilage matrix is disrupted~\cite{nickien2013changes}, or if there is natural collagen network disruption~\cite{thambyah2007degeneration} as in the case with early OA. These findings suggest a significant opportunity for cartilage health assessment in clinical settings, if morphology under load can be observed non-destructively.

A solution to these identified challenges may lie in optical coherence tomography (OCT)~\cite{drexler2008optical} and its functional extension, polarisation-sensitive OCT (PS-OCT)~\cite{de2017polarization}. Already established for assessing cartilage degeneration in unloaded conditions, OCT and PS-OCT offer a powerful, non-invasive method for studying cartilage~\cite{drexler2001correlation,chu2007clinical,chu2004arthroscopic,brill20153d,brill2015optical,li2005high,herrmann1999high,zhou2020detecting}. OCT, the optical analogue of ultrasound, operates on the principle of low-coherence interferometry, while PS-OCT, a functional extension of this technology, enables the measurement of birefringence and optical axis, optical properties linked to collagen organisation and orientation respectively. 

One of the earliest notable applications of OCT to assess cartilage health came from Chu \textit{et al.}~\cite{chu2004arthroscopic}, who used an arthroscopic OCT probe to assess cartilage surface quality \emph{in vivo}, showing strong correlation with histopathology. Brill \textit{et al.}~\cite{brill20153d,brill2015optical} aimed at improving surface quality assessment with OCT but ultimately struggled to reliably distinguish between different stages of degeneration. The problem was clear: surface roughness alone was not enough; a new approach was needed to capture underlying changes in collagen structure that precede surface morphological damage.

PS-OCT emerged as a clear contender for this, promising deeper structural insights by exploiting cartilage birefringence. Drexler \textit{et al.}~\cite{drexler2001correlation} was the first to demonstrate that birefringence patterns in PS-OCT images corresponded with cartilage health, their disappearance marking early degeneration. While some studies supported this~\cite{chu2007clinical,li2005high}, others found inconsistencies~\cite{ugryumova2005collagen,xie2006determination,xie2008topographical}, casting doubt on PS-OCT reliability as a stand-alone diagnostic tool. The flaw, however, was likely not in the technology but possibly in the analysis, early studies relied on qualitative interpretation of birefringence patterns, making them vulnerable to bias. Recognizing this limitation, Shyu \textit{et al.}~\cite{shyu2009diagnosis} introduced the birefringence coefficient (BRC), a metric dependent on the gradient of birefringence with depth and showed a strong correlation of BRC to degeneration using histopathology. Brill \textit{et al.}~\cite{brill2016polarization} further analysed BRC and birefringence banding patterns but reported weaker correlations with degeneration, though their findings were constrained by sample size. Zhou \textit{et al.}~\cite{zhou2020detecting} refined BRC analysis by accounting for zonal variations within cartilage, further enhancing the potential of PS-OCT for early OA detection, though their findings were again ultimately based on a limited patient cohort. More recently, Martin \textit{et al.}~\cite{martin2022minimally} conducted a feasibility study evaluating PS-OCT for \emph{in vivo} cartilage degeneration assessment using an endocatheter. Their results demonstrated the promise of PS-OCT as a minimally invasive tool for OA prediction, however its predictive value remains inconclusive.

Despite these advances, one problem remains: all these studies have been conducted on unloaded tissue. Static imaging, while valuable, offers only a partial picture. Cartilage is a dynamic tissue, and its response to mechanical stress is fundamental to both its function and its failure. To bridge the gap between structure and function, PS-OCT must be extended to capture cartilage under load, unlocking new possibilities for physiologically relevant assessment of cartilage health and function.

Recent efforts by the present authors have sought to address this gap by integrating PS-OCT with mechanical compression~\cite{goodwin2018quantifying,goodwin2021impact,goodwin2022detection}. Initial work has demonstrated the feasibility of using PS-OCT to quantify birefringence changes in loaded cartilage, showcasing strong correlation between birefringence response to load and cartilage health. However, conventional PS-OCT suffers from a fundamental limitation: birefringence accumulates along the light path, yielding a depth-integrated measurement rather than a localised assessment of tissue birefringence. This results in a masking, where important structural features are obscured~\cite{hitzenberger2001measurement,fan2013imaging,li2018robust,villiger2018optic}. To address the limitations of conventional PS-OCT, depth-resolved PS-OCT has been developed~\cite{fan2013imaging,li2018robust,ju2013advanced}, enabling localised measurements of birefringence and the optical axis at different depths within the tissue. This technique overcomes the cumulative effects that complicate the interpretation of traditional PS-OCT data, offering a clearer and more accurate assessment of collagen organisation.  

The primary aim of this study is to evaluate the capability of depth-resolved PS-OCT to study the microstructural response of cartilage to compressive loading. To achieve this, we employ depth-resolved PS-OCT to examine cartilage chemically fixed under load and directly compare our findings with DIC microscopy, effectively benchmarking our approach against Thambyah’s methodology~\cite{thambyah2006micro}. While this study does not yet implement dynamic imaging, our findings suggest that depth-resolved PS-OCT has the potential to address fundamental challenges in cartilage mechanics research, particularly the lack of non-destructive techniques for assessing cartilage under load and the absence of dynamic imaging for real-time mechanical response. This approach has significant implications, not only for the early detection of osteoarthritis (OA) but also for assessing disease progression and evaluating the effectiveness of emerging cartilage therapeutics. 

\section{Theory and computational methods}
\subsection{Depth-resolved PS-OCT}

Many biological tissues are made up of proteins like collagen that have a highly organised and well-aligned fibrillar structure~\cite{ghosh2011tissue}, which causes the refractive index to vary with the polarisation of light. This optical anisotropy is known as birefringence and high birefringence is an indicator of high organisation and alignment.

\begin{figure}[tb!]
    \centering
    \includegraphics[width=1\linewidth]{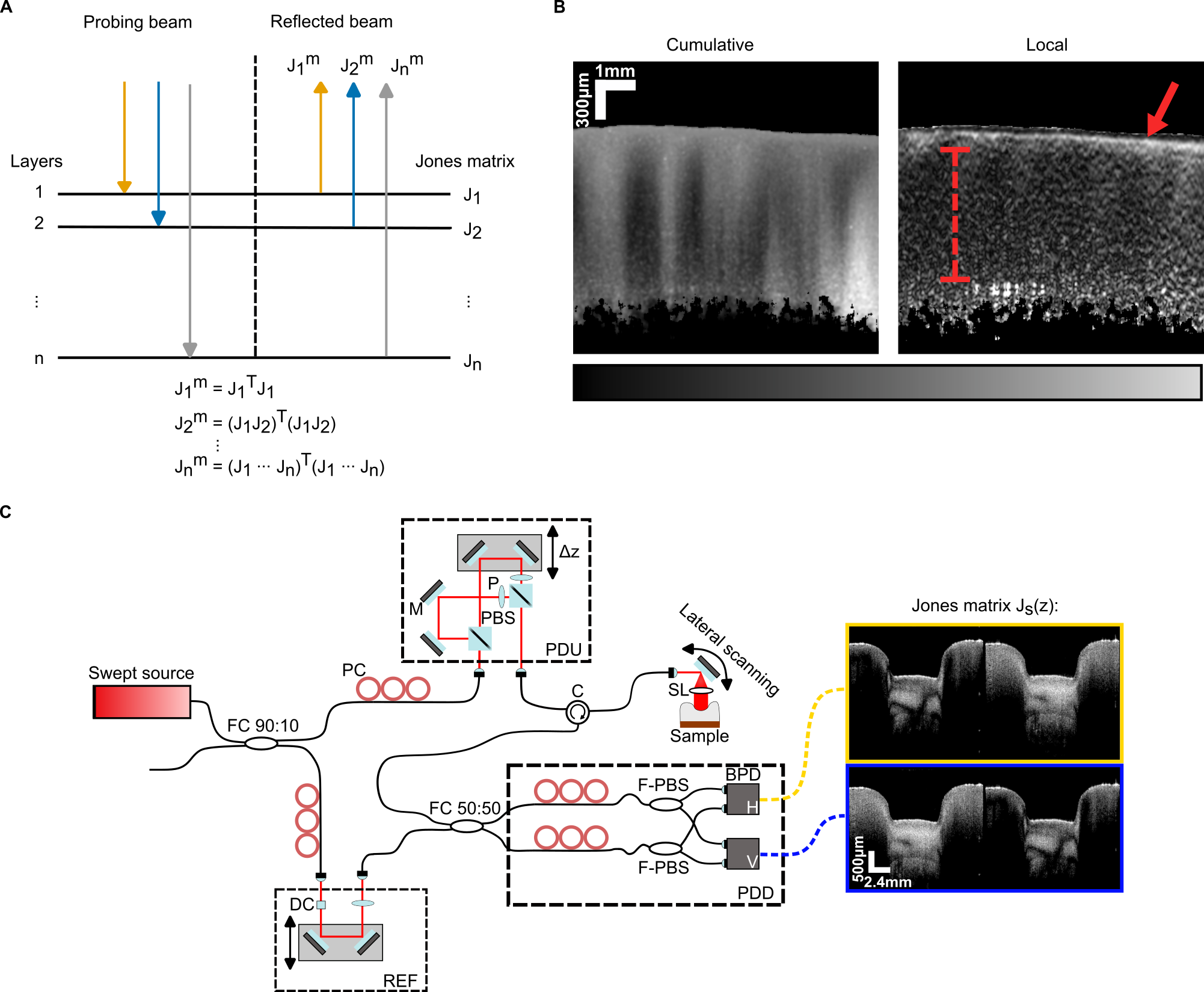}
    \caption{(A) Diagram showcasing the phenomenon of birefringence accumulation. $\mbox{J}_{i}^{m}$ is the measured Jones matrix and $\mbox{J}_{i}$ is the true `local’ matrix for the layer. We denote the reverse transmission Jones matrix with its transpose for ease of representation instead of its proper form diag(1,-1)·J$^{\mbox{T}}$·diag(1,-1)~\cite{gil2022polarized}. (B) (left) Cumulative birefringence image. (B)(right) depth-resolved birefringence image, with a red arrow pointing towards a highly birefringent tangential layer and a dashed red line indicating a weakly birefringent layer beneath. The colour bar goes from 0 to $\pi$ for the cumulative image and 0 to $2.2 \times 10^{-3}$ for the local image. (C) (left) Schematic representation of the optics within the custom-built PC-OCT system adapted from our previous work~\cite{tharan2025towards}. (C) (right) Intensity representation of the Jones matrix of bovine cartilage acquired by the system. Abbreviations: PDU - Polarisation Delay Unit; P - Polariser; REF - Reference Arm; PDD - Polarisation Diversity Detection; FC - Fibre Coupler; C - Circulator; PC - Polarisation Controller; M - Mirror; PBS - Polarisation beam splitter; F-PBS - Fibre based PBS; SL - Scan lens; DC - Dispersion compensation block for scan lens; BPD - Balanced photodetector; H - Photodetector for horizontal polarisation state; V- Photodetector for vertical polarisation state; $\mbox{J}_{\mbox{s}}(\mbox{z})$ - Jones Matrix; $\Delta$z - Induced depth delay.}
    \label{Theory:System}
\end{figure}

The Jones matrix of a sample encapsulates how it transforms the polarisation state of light as it passes through, representing both the birefringence strength and the optical axis of the sample. Measuring local or depth-resolved polarimetric properties may be done using a PS-OCT system that directly measures the sample Jones matrix~\cite{villiger2018optic,villiger2016deep,ju2013advanced}. The Jones matrices are still cumulative (Fig.~\ref{Theory:System}A), but various correction methods allow for the extraction of depth-resolved parameters from the Jones matrix~\cite{li2018robust,fan2013imaging}, which significantly simplifies data interpretation and can reveal structures previously obscured by cumulative birefringence effects. Figure~\ref{Theory:System}B presents a comparison between cumulative and local birefringence from healthy cartilage. The depth-resolved image distinctly shows two layers: a highly birefringent tangential layer (indicated by the red arrow) and a weakly birefringent layer beneath it (indicated by the dashed red line). Such clear differentiation is not apparent in the cumulative image, which only suggests the presence of birefringence without clear stratification.

Measurement of the tissue Jones matrix involves probing the sample with two distinct polarisation states of light. This can be achieved either simultaneously~\cite{villiger2016deep,ju2013advanced} or sequentially~\cite{fan2012full}. In simultaneous probing, both polarisation states are incident onto the sample at the same time, but a phase modulation is applied to one of the states so that it appears deeper in depth within the final OCT image. This spatial separation in depth allows the two states to be distinguished. Alternatively, sequential probing can be implemented using an electro-optic or acousto-optic modulator to rapidly switch between the two polarisation states, thereby separating them in time rather than space. Regardless of the method used to deliver the two polarisation states, a typical PS-OCT system employs two detection channels to independently measure the horizontally and vertically polarised components of the backscattered light from the sample. In the case of simultaneous probing, this results in 2 image sets: the first pair (H1 and V1) corresponds to the undelayed polarisation state, while the second pair (H2 and V2) correspond to the delayed state, appearing deeper in the OCT image due to the imposed delay (Fig.~\ref{Theory:System}C). In this work a simultaneous probing scheme is utilised. 

\subsection{Data processing}
\subsubsection{Pre-processing and intensity computation}

The initial image reconstruction process encompasses several critical steps: removal of invalid points, background subtraction, chromatic dispersion compensation, demodulation of the delayed polarisation state, and the generation of the cumulative Jones matrix. Firstly, the removal of invalid points addresses irregularities in the acquired spectra during laser wavelength tuning, a characteristic artifact of the semiconductor swept-source laser used in this study~\cite{bonesi2014akinetic}. Following this, background subtraction eliminates fixed pattern noise by subtracting the mean OCT spectrum. Chromatic dispersion compensation corrects for adverse wavenumber~(k) dependent effects that reduces system depth resolution. During the demodulation stage, the axial alignment of the delayed and non-delayed images is enforced. The delay between the two polarisation states, $\Delta$z, is determined with sub-pixel accuracy by iteratively minimising the cross-correlation between the delayed and non-delayed OCT signals . Subsequently, the sample’s cumulative Jones matrix, $\mbox{J}_{s}$, is formed. This matrix encapsulates the horizontal (H) and vertical (V) components of the non-delayed polarisation state in its first column and the corresponding components of the delayed state in its second column:

\begin{equation}
\mbox{J}_{s}(\mbox{k}) =  
\begin{bmatrix}
\mbox{H1} & \mbox{H2} \\
\mbox{V1} & \mbox{V2} 
\end{bmatrix}
\end{equation}

Intensity images are then derived by computing the determinant of \( \mbox{J}_{s}(\mbox{z}) = \mathrm{FT}[\mbox{J}_{s}(\mbox{k})] \)~\cite{li2014coherent}, where FT represents the Fourier Transform and \( \mbox{J}_{s}(\mbox{z}) \) belongs to the special unitary group \(\mathrm{SU(2)}\).

\subsubsection{Fibre and system distortion correction}

Fibre and system distortions are corrected according to the procedures outlined by Li \textit{et al}~\cite{li2018robust,hackmann2024quantification}. These distortions primarily stem from two sources: (1) polarisation mode dispersion (PMD) due to k-dependent birefringence in optical fibres~\cite{villiger2013artifacts}, and (2) asymmetry in probing and detection paths, which induces circular birefringence and complicates OpAx computation~\cite{li2018robust}.

In summary, transpose symmetry of the Jones matrices is numerically enforced using an \textit{a posteriori} computed k-dependent correction matrix,  \( \mbox{J}_{Sy}(\mbox{k}) \), minimising system-induced circular birefringence. Subsequently, PMD compensation is performed by applying another \textit{a posteriori} estimated correction matrix, \( \mbox{J}_{P}(\mbox{k}) \), determined through a minimisation procedure. The correction matrices are applied to all measured \( \mbox{J}_s \) via matrix multiplication:

\begin{equation}
\mbox{J}_{C} = \mbox{J}_{P}^{T} \cdot (\mbox{J}_{Sy} \cdot \mbox{J}_{S}) \cdot \mbox{J}_{P}
\end{equation}

\noindent where \( \mbox{J}_{C} \) represents the fibre distortion corrected sample Jones matrix.

\subsubsection{Depth-resolved OpAx and birefringence computation}

After transpose symmetry is enforced and PMD correction, SU(2) Jones matrices are converted to SO(3) rotation matrices~\cite{gordon2000pmd} to avoid the computationally intensive coherent averaging procedure usually done in Jones matrix polarimetry~\cite{ju2013advanced,villiger2016deep,braaf2014fiber}. This conversion is achieved by decomposing \( \mbox{J}_{C} \) into its retardation and diattenuation vectors, \( \mbox{r}_n \) and \( \mbox{d}_n \) (\( n = 1, 2, 3 \)):

\begin{equation}
    \mbox{d}_n - i \mbox{r}_n = \frac{\mbox{q}_n \cdot \cosh^{-1} \left( \frac{\mbox{q}_0}{2} \right)}{\sinh \left( \cosh^{-1} \left( \frac{\mbox{q}_0}{2} \right) \right)},
\end{equation}

\noindent where \( \mbox{q}_n = \text{Trace}(\sigma_n \cdot \mbox{J}_C) \), \( \mbox{q}_0 = \text{Trace}(\mbox{J}_C) \), and \( \sigma_n \) is the standard Pauli basis. The \( \mbox{d}_n \) component is ignored while \( \mbox{r}_n \) is used to define an SO(3) rotation matrix via Rodrigues' rotation formula. To simplify the procedure, we neglect diattenuation, the polarisation dependent attenuation of light, which is typically negligible in biological tissue.

Spatial averaging of these SO(3) matrices is performed using a 35~$\mu$m 3D Gaussian moving window to minimise speckle effects on OpAx measurements. Depth-resolved rotation matrices, \( \mbox{R}(\mbox{z}) \), are then computed layer by layer using a recursive peeling procedure described by Li \emph{et al.}~\cite{li2018robust}. Given \( \mbox{R}(\mbox{z}) \) as:
\[
\mbox{R(z)} = \begin{bmatrix}
\mbox{R}_{11} & \mbox{R}_{12} & \mbox{R}_{13} \\
\mbox{R}_{21} & \mbox{R}_{22} & \mbox{R}_{23} \\
\mbox{R}_{31} & \mbox{R}_{32} & \mbox{R}_{33} 
\end{bmatrix},
\]

\noindent the depth-resolved OpAx, \( \theta(\mbox{z}) \), is computed using:
\begin{equation}
    2\theta = \text{atan2} \left( \frac{\mbox{R}_{32} - \mbox{R}_{23}}{\mbox{R}_{13} - \mbox{R}_{31}} \right),
\end{equation}

\noindent where \(\text{atan2}\) is the four-quadrant arctangent function. Further spatial averaging with a 50~$\mu$m axial window is performed to reduce noise. The measured OpAx is shifted by an unknown amount, \(\phi\), due to birefringence induced by sample arm fibres~\cite{li2018robust}. Further, there is an inherent ambiguity in the sign of \( \theta(\mbox{z}) \)~\cite{park2005optic}. Both the \(\phi\) shift and sign ambiguity are calibrated using a phantom with a known OpAx orientation, as described later. 

The local phase retardation, \( \delta \), is also computed from R(z):
\begin{equation}
    \delta = \sqrt{(\mbox{R}_{32} - \mbox{R}_{23})^2 + (\mbox{R}_{13} - \mbox{R}_{31})^2},
\end{equation}
\noindent from which the local birefringence, \( \Delta n \), is extracted:
\begin{equation}
    \Delta n = \frac{\delta}{2\mbox{k}_c\Delta \mbox{z}},
\end{equation}
\noindent where \( \mbox{k}_c \) is the central wavenumber of our laser source and \(\Delta \mbox{z}\) is axial thickness of a pixel.

\subsubsection{Depolarisation index computation}

Depolarisation refers to the reduction of the degree of polarisation of light as it passes through a medium. Depolarisation is typically caused by diattenuation or regions of high scattering, which indicate significant sample inhomogeneity. 

Since the sample is probed with two distinct polarisation states, we are able to measure the unambiguous depolarisation index (DOP). This index is independent of the input polarisation state and is purely a property of the sample~\cite{lippok2015degree}. We compute the DOP by first converting $\mbox{J}_c$ into an SU(4) Mueller matrix:

\begin{equation}
    \mbox{M(z)} = \mbox{A}(\mbox{J}_{C} \otimes \mbox{J}_{C})\mbox{A}^{-1} ,
\end{equation}

\begin{equation}
    \mbox{A} =  \begin{bmatrix}
1 & 0 & 0 & 1 \\
1 & 0 & 0 & -1 \\
0 & 1 & 1 & 0 \\
0 & i & -i & 0 
\end{bmatrix} , 
\end{equation}

\noindent where $\otimes$ is the standard tensor product. Next, spatial averaging is done on M(z) by the same kernel used for the SO(3) matrix. Averaging the Mueller matrices is a necessary step in computing the DOP or else M(z) remains non-depolarising. We refer to Lippok \emph{et al}.~\cite{lippok2015degree} for an extended description. From the averaged M(z), we compute the DOP using the equation:

\begin{equation}
\mbox{DOP} = \sqrt{\frac{\mathrm{Trace}(\mbox{M}^T \mbox{M}) - \mbox{M}_{00}^2}{3 \mbox{M}_{00}^2}},
\end{equation}

\noindent where $\mbox{M}_{00}$ is the first element of M(z). A  DOP value equal to one represents a perfectly non-depolarising region and a DOP value of zero represents a perfectly depolarising region.

\subsubsection{Data presentation}

All birefringence and DOP images presented are overlaid over the intensity images to create an image consisting of both structure and polarimetric function. Similarly, OpAx and birefringence are combined, masking low birefringence regions in black within the OpAx image to avoid unreliable or random optical axis values~\cite{li2018robust}. We use the `ametrine’~\cite{geissbuehler2013display} colour-blind friendly map for birefringence and DOP and the cyclic `romao’~\cite{crameri2020misuse} colour-blind friendly map for OpAx. All PS-OCT images presented in this work undergo a final processing step in which a Gaussian-weighted average is computed over 7 spatially adjacent PS-OCT frames to enhance image quality.

\subsubsection{Tractography}

Tractography is a technique used to visualise and map the pathways of fibrous structures, such as nerve fibres in the brain or collagen fibrils in tissues. This method reconstructs these pathways by tracking the paths of fibres, allowing for detailed visualisation of complex networks.This approach facilitates more accurate comparisons with DIC microscopy. 

For a given \emph{en face} plane, a 2D vector field is extracted from the OpAx as follows:
\begin{equation}
    \mathbf{F}(\mathbf{x}) = \begin{bmatrix}
        \cos(\theta(\mathbf{x})) \\
        \sin(\theta(\mathbf{x}))
    \end{bmatrix}
\end{equation}

\noindent Streamlines, which represent paths tangent to the vector field, are computed by solving the differential equation:
\begin{equation}
    \frac{d\mathbf{r}(t)}{dt} = \mathbf{F}(\mathbf{r}(t))
\end{equation}

\noindent The streamline starting from \(\mathbf{x}\) is found by integrating the vector field both forward and backward:
\begin{equation}
\begin{aligned}
    \mathbf{r}(t; \mathbf{x}) &= \mathbf{x} + \int_{0}^{t} \mathbf{F}(\mathbf{r}(\tau; \mathbf{x})) \, d\tau \quad \text{for} \quad t \geq 0, \\
    \mathbf{r}(t; \mathbf{x}) &= \mathbf{x} - \int_{0}^{-t} \mathbf{F}(\mathbf{r}(\tau; \mathbf{x})) \, d\tau \quad \text{for} \quad t < 0,
\end{aligned}
\end{equation}
where \(t\) is the parameter along the streamline, and \(\tau\) is a dummy integration variable. We use the adaptive-step Dormand-Prince integrator for accurate and stable numerical integration.

To avoid sparse or cluttered regions, we utilise the effective seeding strategy described by Jobard and Lefer~\cite{jobard1997creating} (see supplementary materials). The code developed by Ma~\cite{ma_github} was modified for our specific use. Once the streamlines are generated, they represent the paths of the collagen fibrils within the imaged tissue.

\section{Materials and methods}
\subsection{Imaging instrument}

A custom-built PS-OCT system, designed for measuring depth-resolved parameters, is employed in this study (Fig.~\ref{Theory:System}C). This system utilises a polarisation delay unit (PDU) to passively probe the sample with 2 polarisation states. In doing so, the sample Jones matrix is captured in a single scan. 

The PDU splits the incoming light into orthogonal polarisation states using a polarisation beam splitter (PBS). One of these polarisation states is delayed by a distance \(\Delta\)z relative to the other. After the delay, the two polarisation components are recombined before exiting the PDU. This delay introduces a depth encoding of one of the polarisation states, enabling the system to distinguish between the two states within the OCT image. The system employs two detectors to simultaneously measure the light horizontal (H) and vertical (V) polarisation components. This configuration results in four distinct images: H1, V1, H2, and V2, as described earlier (Fig.~\ref{Theory:System}C). These four images collectively represent the four components of the cumulative Jones matrix of the sample.

A 1310~nm centred Akinetic semiconductor swept laser source (Insight Photonics Ltd, Lafayette, US) is used in this study. This source features a bandwidth of 85~nm, a 100~kHz sweep rate, and an average output power of 54~mW. The system achieves an axial resolution of 10~µm in air, after compensating for chromatic dispersion, and a lateral resolution of 31 µm. The optical power incident on the sample is 8 mW (4 mW per polarisation state). The Akinetic swept source offers numerous advantages for PS-OCT imaging. Its wavelength tuning is achieved without any mechanical motion, resulting in intrinsic phase stability and minimal wavenumber jitter~\cite{bonesi2014akinetic}. This intrinsic stability eliminates the need for jitter correction, thereby simplifying the optical setup~\cite{braaf2011phase}. Additionally, the long coherence length of the Akinetic source (10 cm) removes the need for roll-off correction, enabling accurate PS-OCT reconstruction with depth.

\subsection{OpAx calibration assessment}

We have previously evaluated the PS-OCT system capability to measure the OpAx by imaging a birefringent phantom at various orientations~\cite{tharan2025towards}. PS-OCT scans of the phantom were acquired at orientations ranging from 0° to 180°, in 10° increments. The initial orientation was selected as the reference (0°), and subsequent scans were compared against this baseline. We found excellent agreement between the expected orientation and the measured OpAx and identified a root-mean-square error of 5°. To correct both the OpAx shift, \(\phi\), and sign ambiguity we constructed a calibration phantom by mounting multiple identical phantoms, each aligned at a known orientation. This arrangement enabled simultaneous estimation of \(\phi\) and the sign of the OpAx within a single PS-OCT scan. Provided the optical fibres of the PS-OCT system remain undisturbed, the resulting calibration can be applied to correct subsequent PS-OCT measurements within a reasonable time frame.

\subsection{Sample preparation}

Healthy bovine patellae, known for being an excellent model for OA progression in humans~\cite{hargrave2013bovine}, are obtained from prime animals (5–7 years old), collected shortly after slaughter from a local abattoir and stored at $-18^\circ\text{C}$. The patellae are thawed under cold running water for at least 30~minutes. Carefully selecting a flat region, a cartilage-on-bone block with \textit{en face} dimensions of $15 \times 15$~mm is sawn from the patella, distal-lateral to the knee joint line. The block includes the full cartilage thickness and a minimum 10~mm depth of bone for support under load. The sample is embedded in a stainless-steel holder with dental cement before equilibration in 0.15~M saline for one hour prior to loading.

\subsection{PS-OCT comparison with DIC microscopy}

\begin{figure}[b!]
    \centering
    \includegraphics[width=1\linewidth]{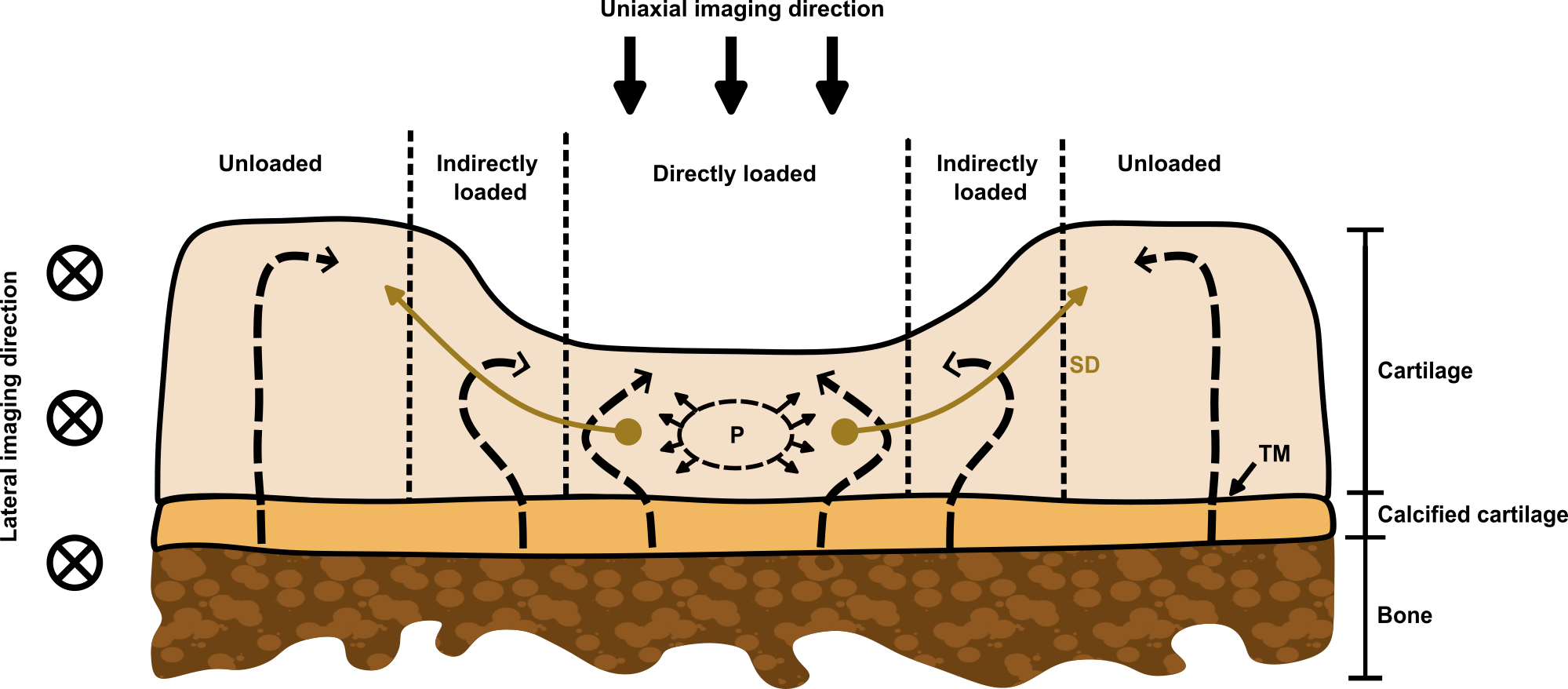}
    \caption{Schematic illustrating the typical collagen fibril deformation pattern in cartilage under indentation. The three regions of analysis are clearly marked, along with the imaging directions. Collagen fibrils are represented by dashed black arrows, while the shear discontinuity is indicated by a solid arrow labelled  SD. The build-up of hydrostatic pressure is shown by the dashed oval labelled  P, with surrounding arrows depicting fluid movement. Abbreviations: P - Hydrostatic pressure; TM – Tidemark}
    \label{fig:methods:regionsofinterest}
\end{figure}

Loading is performed using a 5~mm stainless steel indenter with an Instron universal materials testing machine (\textit{Illinois Tool Works} Inc, Massachusetts, US). Samples are submerged in a 0.15 M saline bath, and a stress of approximately 4 MPa is applied for one hour to facilitate creep loading. Post-loading, samples are formalin-fixed in their indented state for twelve hours at room temperature, followed by decalcification over three days for easier sectioning~\cite{thambyah2006micro}. A diametric cut through the decalcified sample exposed the cross-sectional profile of the loaded matrix and its continuum. The tissue is examined under a DIC optical microscope, using chondrocytes as markers to assess fibrillar orientation. While DIC microscopy cannot directly observe collagen fibrils, it has been shown that fibril orientation aligns with the long axis of chondrocytes~\cite{kaab2000effect}. Thus, by observing the continuity lines formed by the chondrocytes, the overall fibril orientation can be inferred~\cite{thambyah2006micro}.

Following DIC microscopy, the tissue is imaged using PS-OCT in two orthogonal directions. The first is the uniaxial direction where the scanning head is orthogonal to the articular cartilage surface. The second is the lateral direction where the PS-OCT scanning head is orthogonal to the sagittal plane, offering the full cartilage-on-bone side view. The rationale for scanning in two directions is to understand the anisotropy-related responses, as well as to evaluate the comparative efficacy of scanning one direction versus the other. All PS-OCT images presented in this work include insets indicating the scan direction, where the axial direction is indicated by: (\( \downarrow \)), and the lateral direction is indicated by: (\( \otimes \)). A total of five samples are analysed using the same experimental protocol. The results shown in this paper are from a representative sample.

The PS-OCT system performs line scans (A-scans) into the tissue, which are compiled to reconstruct planar views of cartilage structure. In the uniaxial imaging direction, line scans are directed perpendicular to the cartilage surface and lateral translation across the surface generates a series of cross-sectional B-scans (sagittal plane image). In the lateral imaging direction, line scans are directed perpendicular to the exposed cartilage-on-bone profile after sectioning and laterally scanned across. This similarly yields sagittal cross-sectional images, and stacking these B-scans enables reconstruction of transverse plane \textit{en face} images of the cartilage-on-bone view, providing a comparable perspective to the uniaxial approach. Important to note is that the resolution of the transverse plane lateral PS-OCT images (30~\(\mu\)m by 30~\(\mu\)m) is not the same as the sagittal plane uniaxial images (30~\(\mu\)m by 7~\(\mu\)m).



The indented cartilage can be divided into three distinct regions for analysis as indicated in Figure~\ref{fig:methods:regionsofinterest}: (1) the directly loaded region, where the cartilage is in direct contact with the indenter; (2) the indirectly loaded region, where the non-contact surface of the cartilage is pulled inward toward the centre of indentation; and (3) the unloaded region, where the influence of the loading is minimal and the cartilage surface retains its typical morphology.

\section{Results}

\begin{figure}[p!] 
    \centering
    \includegraphics[width=1\linewidth]{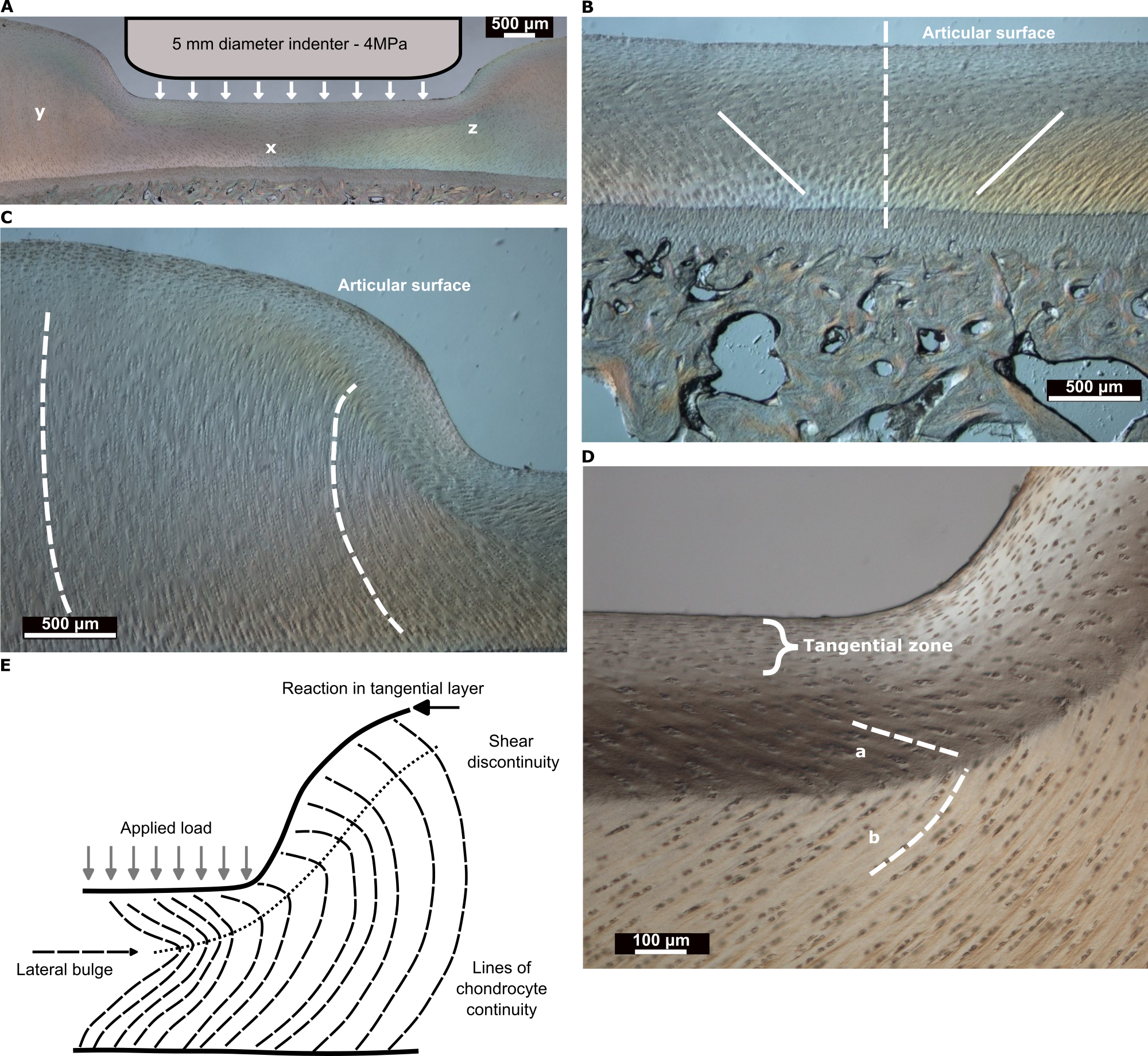}
    \caption{(A) Overall deformation profile of the fixed cartilage tissue, where the indentation span is shown. (B) Shows region x in (A) at higher magnification. The dashed line is the axis of which on either side there is a seemingly equal and opposite `shearing’ as indicated by the alignment of the chondrocytes (white lines). (C) High magnification of region y in (A) comparing the chondrocytes alignment (dotted lines) in two regions of the non-directly loaded cartilage. (D) Shows region z from (A) at higher magnification. The tangential zone is easily discerned from the chondrocyte cell alignment. The dashed lines, shaped like a chevron, represent the alignment of chondrocytes. The lines present an intense shearing, as the cells without any loading would have been aligned radially with respect to the surface. Note also the change in colour and contrast between regions a and b. This boundary coincides with the apex of the chevron indicating the shear discontinuity. (E) A schematic redrawn from Thambyah \textit{et al.}~\cite{thambyah2006micro}, describing the mechanical basis for the chondrocyte alignment in cartilage under load, and the formation of the shear discontinuity.}
    \label{fig:results:DIC}
\end{figure}

Following creep loading and formalin-fixing under load, samples display a fixed deformation that closely followed the footprint of the indenter (Fig.~\ref{fig:results:DIC}A). As reported previously~\cite{thambyah2006micro}, similar patterns of chondrocyte alignment are observed, and, hence, only some of the more relevant observations are reported here. The loading pattern is `axisymmetric’ (Fig.~\ref{fig:results:DIC}B) such that the cartilage deformed in a relatively uniform way on either side of the centre of loading. The dissipation of stress into the wider cartilage continuum is evident by the gradual transition of a curvilinear chondrocyte alignment to a more radial (unloaded) configuration (Fig.~\ref{fig:results:DIC}C). Using DIC microscopy, the pattern of chondrocytes show shearing in the matrix (lines of chondrocyte continuity shown as dotted lines in Fig.~\ref{fig:results:DIC}D), in the form of a `chevron’ pattern.  At the centre of the chevron is a distinct refractile boundary coinciding with the directional change in the chondrocytes. This boundary is the shear discontinuity described and discussed in the introduction. The SD is noted to be a little below 400~$\mu$m from the surface. A schematic showcasing the formation of the SD and associated chondrocyte lines of continuity is presented in figure.~\ref{fig:results:DIC}E.

\begin{figure}[tb!]
    \centering
    \includegraphics[width=1\linewidth]{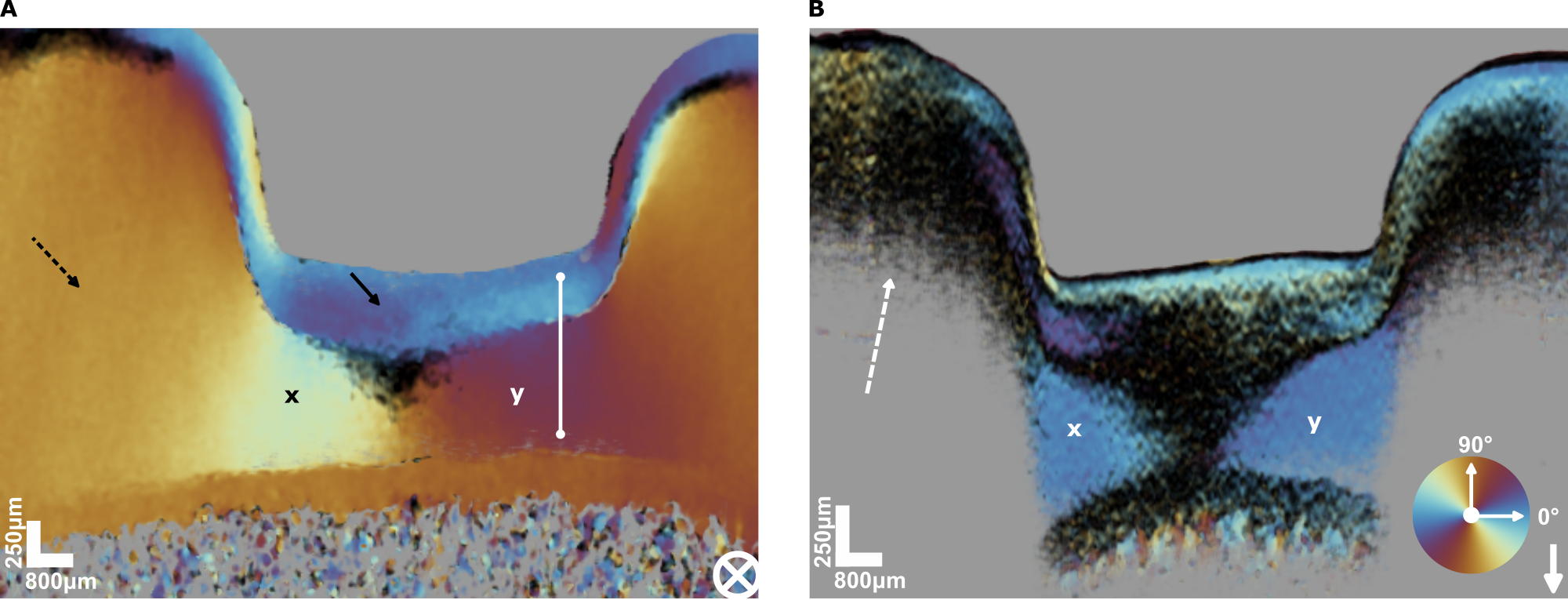}
    \caption{(A) Laterally scanned optical axis image. The black arrows highlight 2 distinct regions of fibrillar orientation, a radial orientation indicated by the orange tones and a tangential  orientation indicated by the blue tones. x and y indicate regions of collagen fibril shear into opposite directions. (B) Uniaxial OpAx image with colour-wheel at bottom right indicating axis orientation. x and y indicate regions where  the OpAx is pointing away from the axisymmetric centre. The white arrow is discussed in the text.}
    \label{fig:results:opax}
\end{figure}

The lateral optical axis (OpAx) image (Fig.~\ref{fig:results:opax}A) of the sample reveals details of the collagen network alignment. The blue-coloured surface layer (Fig.~\ref{fig:results:opax}A, solid black arrow) represents a tangentially directed (0°) network, compared with the approximately radial (90°) orientation of the network as indicated by the orange tones (Fig.~\ref{fig:results:opax}A, dashed black arrow). Also present is a central region in the image that is black. There is also a region indicating intense shear (Fig.~\ref{fig:results:opax}A, labelled as x) where the optical axis is greater than 120°. Across the axisymmetric centre of loading, region y (Fig.~\ref{fig:results:opax}A) has an optical axis of around 45°, which is 90° less than region x. Further, there is  a changing OpAx orientation with depth with colours going from light blue to maroon (Fig.~\ref{fig:results:opax}A, white line), indicating a shift of approximately 60° in the OpAx, representing the `chevron’ patterned shearing of the matrix. In comparison, the uniaxial OpAx image shows the axisymmetric shear (Fig.~\ref{fig:results:opax}B, regions x and y) appearing as the same colour (same optical axis) because of their similar projected orientation relative to the probing direction. The central region of the uniaxial OpAx image appears black, but this region is more widespread than that in the lateral OpAx image. Note that there is a limited level of depth that could be imaged (Fig.~\ref{fig:results:opax}B, dotted arrow). 

\begin{figure}[tb!]
    \centering
    \includegraphics[width=1\linewidth]{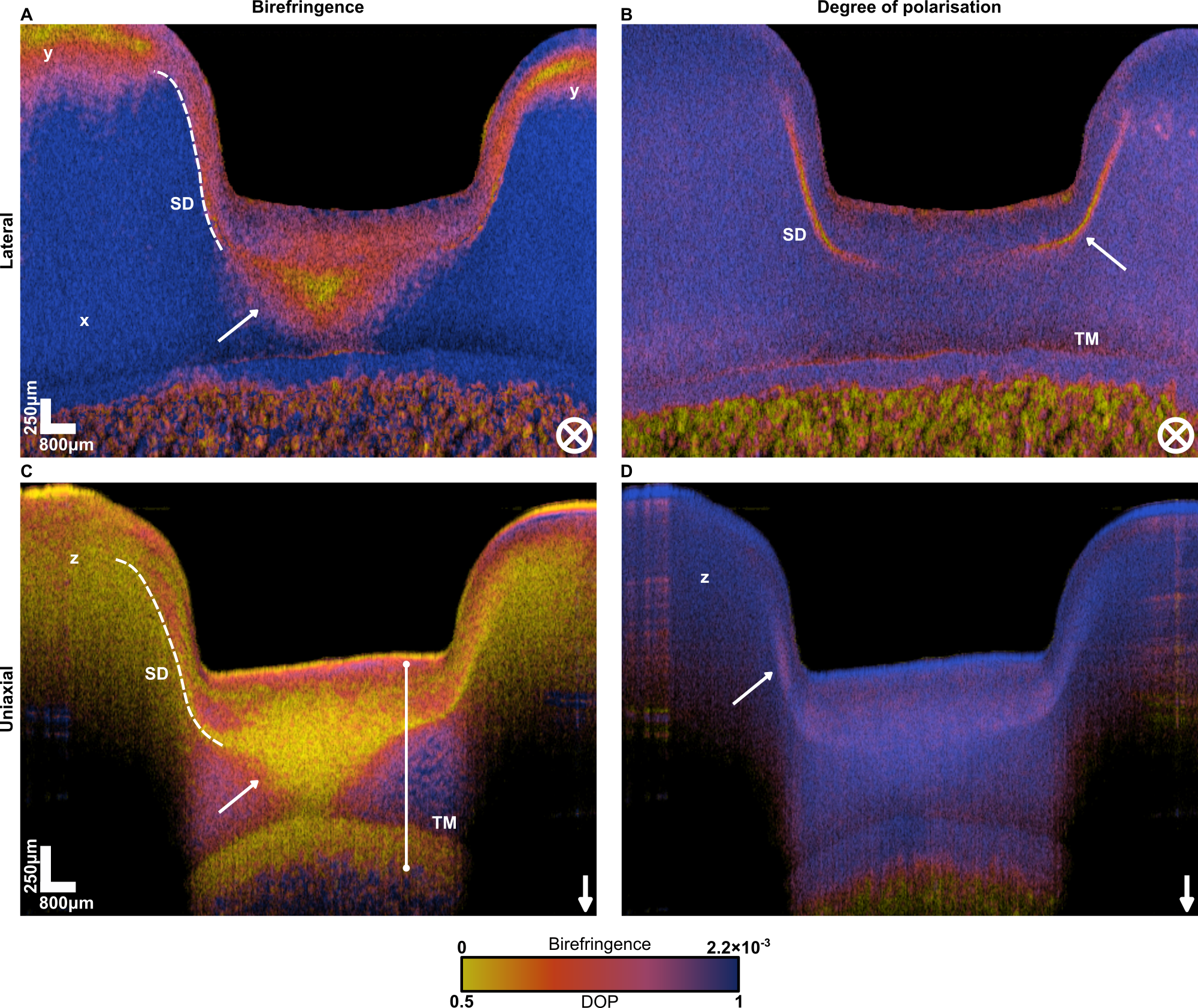}
    \caption{(A, B) Laterally scanned birefringence and DOP images, respectively. (C, D) Uniaxially scanned birefringence and DOP images, respectively. In (A) and (C) the dashed lines refer to the shear discontinuity. The white vertical bar in (C) shows changing birefringence levels in the loaded region with depth, discussed further in the text. The arrows in (A) and (C) refer to a region of low birefringence values at and around the axisymmetric centre, presently denoted as the `dead-zone’ as described in the text. The arrows in (B) and (D) refer to the SD. Regions x, y and z are further discussed in the text. Note that the birefringence values in (A) have been truncated to allow for easy visual comparison between uniaxial and lateral range of birefringence values.}
    \label{fig:results:dnDOP}
\end{figure}

Further insights are gained from the computation of the birefringence and DOP images. In the lateral-scanned images, there is a positive correlation of high birefringence and high DOP (Fig.~\ref{fig:results:dnDOP}A and \ref{fig:results:dnDOP}B, e.g. region x), together with some regions where the correlation is negative (Fig~\ref{fig:results:dnDOP}A and \ref{fig:results:dnDOP}B, regions above the SD). The shear discontinuity (Fig.~\ref{fig:results:dnDOP}A and \ref{fig:results:dnDOP}B, labelled SD), compared to the surrounding, appeared as a `streak’ of both low birefringence and DOP (Fig.~\ref{fig:results:dnDOP}A and \ref{fig:results:dnDOP}B). The decrease in DOP within the SD streak is most prominent within the indirectly loaded region (Fig.~\ref{fig:results:dnDOP}B, white arrow). This SD line extends from either side of the directly loaded centre into the indirectly loaded regions. In the lateral birefringence image, these streaks lead to a localised area of weak birefringence (Fig.~\ref{fig:results:dnDOP}A, labelled as y) within the unloaded cartilage. Discernable in the laterally-scanned images, is the central region where the birefringence drop is intense (Fig.~\ref{fig:results:dnDOP}A, white arrow) while still maintaining a high DOP (Fig.~\ref{fig:results:dnDOP}B). This central region of low to moderate birefringence (yellow to orange hues), presently referred to as the central `dead-zone’, is indicated by the solid white arrow, with the SD streaks continuous from it. Deep below the surface, there is a region of high birefringence and DOP (Fig.~\ref{fig:results:dnDOP}A and \ref{fig:results:dnDOP}B, below TM) corresponding to a region where the optical axis orientation is vertically aligned (Fig.~\ref{fig:results:opax}A, orange hues at the bottom of the image). This region corresponds to the zone of calcified cartilage described previously in the introduction. 

In the uniaxial images, the overall relationship between birefringence and DOP is opposite (Fig.~\ref{fig:results:dnDOP}C and \ref{fig:results:dnDOP}D, compare low birefringence and high DOP in regions Z). This opposite level of intensity is relatively consistent throughout the sample images aside from the regions immediately around the low birefringence centre. Further, there are changing birefringence levels in the loaded region showing high-low-high-low from the cartilage surface, transition zone, the mid-to-deep zones, and calcified cartilage respectively (Fig.~\ref{fig:results:dnDOP}C, vertical line). The SD streaks in the uniaxial birefringence image are clearly discernable (Fig.~\ref{fig:results:dnDOP}C, dashed white line labelled as SD), and in a similar position to that in the lateral image. A decrease in DOP along the SD streak is not noted in the uniaxially-scanned image aside from a slight decrease along the left side of the axisymmetric centre (Fig.~\ref{fig:results:dnDOP}D, white arrow). The birefringence is lower in the streak associated with the SD line imaged from the uniaxial direction compared to that imaged in the lateral direction. In the uniaxial birefringence image, the dead-zone is also visible (Fig.~\ref{fig:results:dnDOP}C, white arrow); however, unlike in the lateral probing direction where it presents as a region of moderate to low birefringence (orange to yellow hues), it appears solely as a region of low birefringence (primarily yellow hues). High DOP is still observed within the dead-zone for the uniaxially scanned DOP image (Fig.~\ref{fig:results:dnDOP}D). The zone of calcified cartilage appears as a region of minimal birefringence but high DOP in the uniaxially scanned images (Fig.~\ref{fig:results:dnDOP}C and \ref{fig:results:dnDOP}D, below TM). 

\begin{figure}[tb!]
    \centering
    \includegraphics[width=1\linewidth]{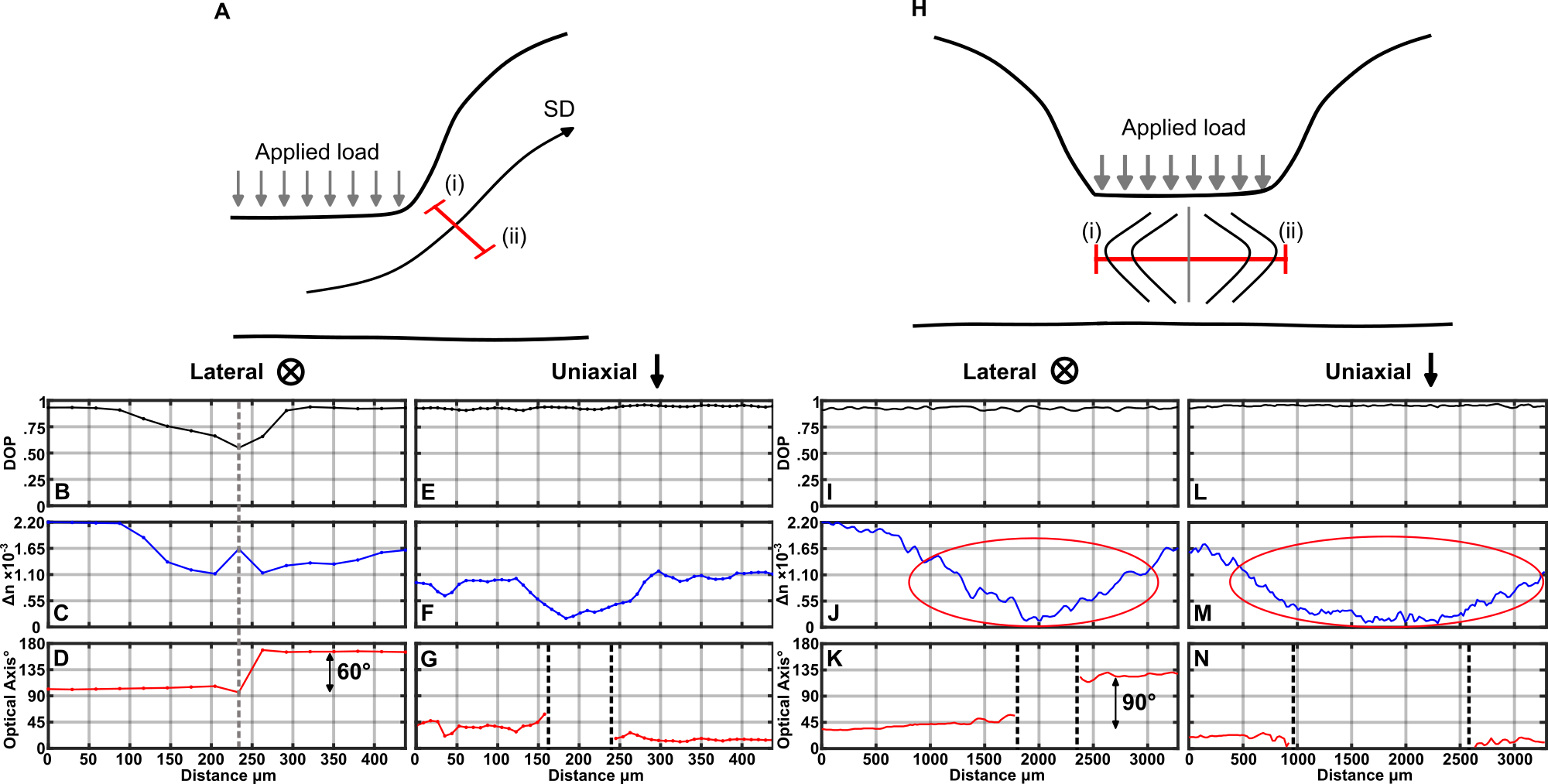}
    \caption{(A) Schematic illustrating the region across the SD from which plots are generated, indicated by the red line. Panels (B, E) show the DOP, (C, F) show the birefringence, and (D, G) show the OpAx across the SD streak, extracted along the red line from (i) to (ii) in (A). (H) Schematic illustrating the region across the dead-zone from which plots are generated, indicated by the red line. Panels (I, L) show the DOP, (J, M) show the birefringence, and (K, N) show the OpAx across the dead-zone, extracted along the red line from (i) to (ii) in (H). The red ovals are discussed in the text.}
    \label{fig:results:plots}
\end{figure}

Plots of the birefringence, OpAx and DOP values on either side of the SD, taken near the edge of the directly loaded region (Fig.~\ref{fig:results:plots}A, red line), are shown in figure~\ref{fig:results:plots}B to \ref{fig:results:plots}G. From the SD streaks in the laterally scanned images (Fig.~\ref{fig:results:opax}A, Fig.~\ref{fig:results:dnDOP}A and \ref{fig:results:dnDOP}B), emerging from either side there is a gradual increase in both birefringence and DOP (Fig.~\ref{fig:results:plots}B, and \ref{fig:results:plots}C, dashed vertical grey line). Interestingly, at the very centre of the SD (dashed vertical grey line) where the DOP reaches its lowest point, there is a sudden spike in birefringence. This spike coincided with the transition of the optical axis between the two shearing orientations (Fig.~\ref{fig:results:plots}D) within the loaded cartilage region (Fig.~\ref{fig:results:opax}A, white line and Fig.~\ref{fig:results:DIC}D), and this transition of the OpAx is measured to be approximately 60°. Similarly for the uniaxially scanned birefringence plot (Fig.~\ref{fig:results:plots}F), emerging from either side of the SD (dashed vertical gray line), a gradual increase in birefringence is observed but with minimal change in DOP (Fig.~\ref{fig:results:plots}E) or optical axis (Fig.~\ref{fig:results:plots}G). Data points where the birefringence is too weak for reliable optical axis measures are removed (e.g. Fig.~\ref{fig:results:plots}G region between 2 dashed vertical black lines). 

Plots of the polarimetric values taken from across the dead-zone (Fig.~\ref{fig:results:plots}H, red line) described earlier, are shown in figure~\ref{fig:results:plots}I to \ref{fig:results:plots}N. In the dead-zone, the birefringence decreases gradually and continuously towards the axisymmetric centre of loading (Fig.~\ref{fig:results:plots}J, red oval), within the lateral imaging direction. This gradual change is less sharp and flattens out in the uniaxial direction (Fig.~\ref{fig:results:plots}M, red oval). A 90° shift in the OpAx is noted across the dead-zone when scanned in the lateral direction (Fig.~\ref{fig:results:plots}K), while the OpAx is approximately constant across the dead-zone in the uniaxial scanned direction (Fig.~\ref{fig:results:plots}N). DOP values are constant and close to unity across the dead-zone for both scan directions (Fig.~\ref{fig:results:plots}I and ~\ref{fig:results:plots}L).

\begin{figure}[tb!]
    \centering
    \includegraphics[width=1\linewidth]{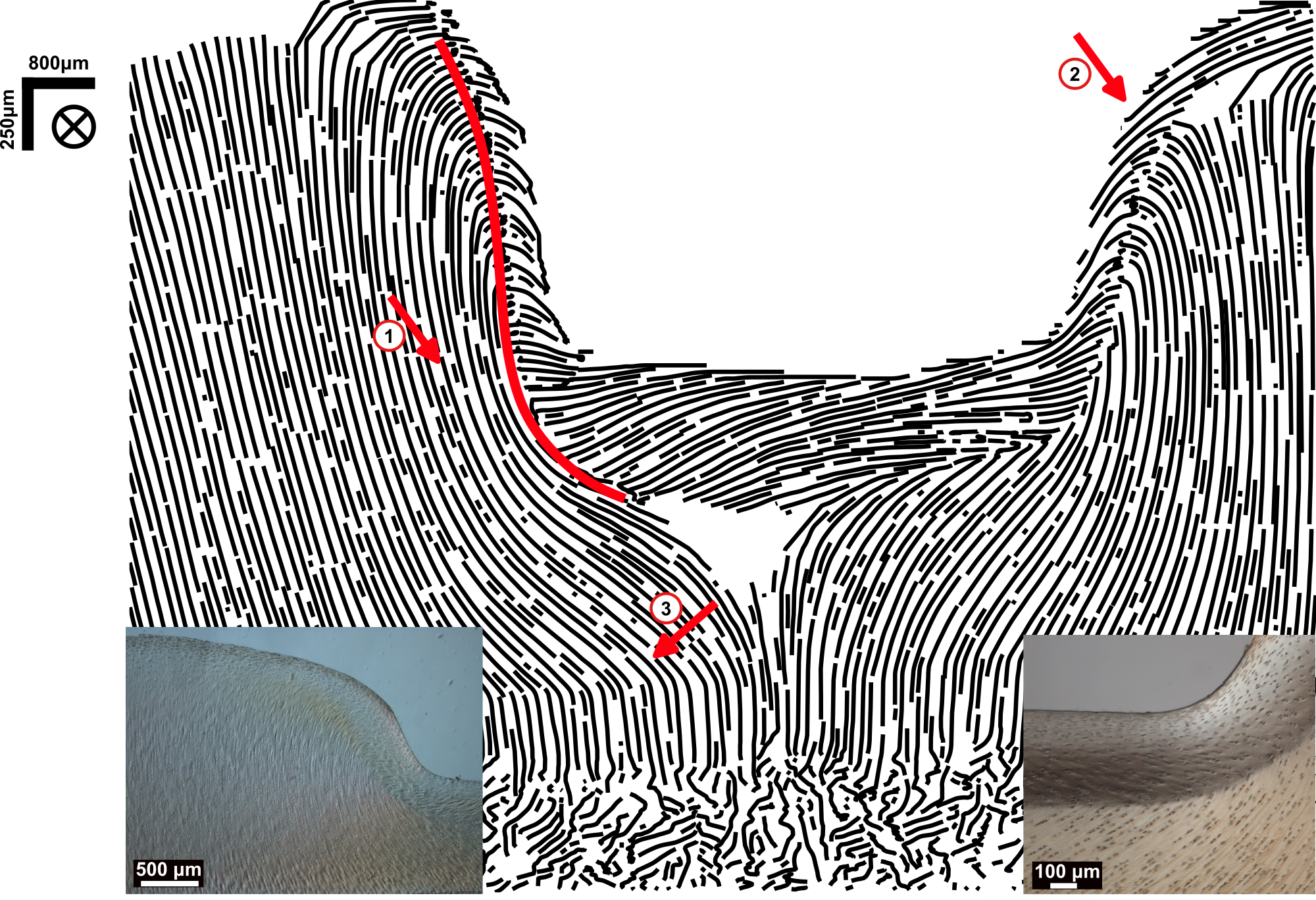}
    \caption{Tractograph computed from the lateral OpAx image (Fig.~\ref{fig:results:opax}A), showcasing a map of the collagen fibrillar orientation. The red arrows labelled 1-3 and the solid red line are discussed in the text. For ease of comparison, insets of the DIC images from Fig.~\ref{fig:results:DIC}C and \ref{fig:results:DIC}D are included.}
    \label{fig:results:tracto}
\end{figure}

Using tractography, the optical axis data for the laterally scanned image (Fig.~\ref{fig:results:opax}A), is plotted as a schematic (Fig.~\ref{fig:results:tracto}). This schematic is taken to imply the overall collagen direction. Three main observations are made. Firstly, the collagen orientation shows evidence of shear, transitioning from the directly loaded region into the adjacent indirectly loaded region (Fig.~\ref{fig:results:tracto}, red arrow (1)). Secondly, the collagen fibrils in the surface layer of the indirectly loaded region exhibit curvilinear patterns (Fig.~\ref{fig:results:tracto}, arrow (2)). Also noteworthy is the directional change of the collagen lines of action when crossing from one side of the SD to the other (Fig.~\ref{fig:results:tracto}, red line). The collagen orientation highlight a sharp transition in fibril alignment across the tidemark (TM), demarcating the boundary between the cartilage and the zone of calcified cartilage (Fig.~\ref{fig:results:tracto}, arrow (3)). Below the TM, the lines of action remain predominantly vertical, reflecting the high shear resistance of the calcified cartilage. Just above the TM, the fibrils shift abruptly, indicating a region of intense lateral shear against the TM.

\begin{figure}[tb!]
    \centering
    \includegraphics[width=1\linewidth]{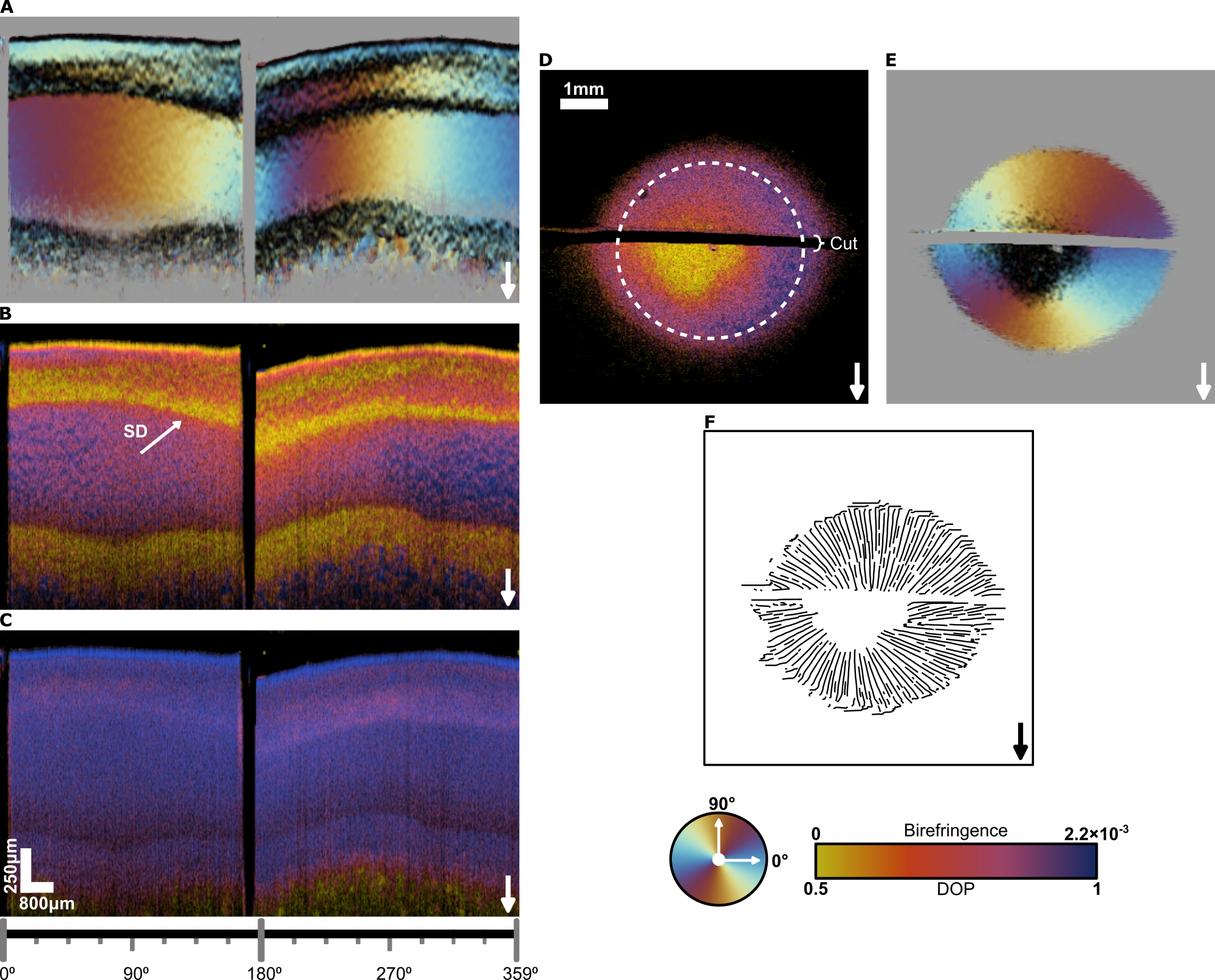}
    \caption{Further analysis into the plane of the image, along the boundary (dotted line) shown in (D), presents an optical axis (A), birefringence (B) and DOP (C) angular view of the depth of the compressed cartilage at a radius of approximately 2~mm from the centre. The method of generating such images is discussed in the supplementary material. There is a clear symmetry of the signal distribution along the full 360° compilation. The distinct contrasts in signal intensity are obvious (B, white arrow), especially coincident to the earlier described SD streak. Sagittal uniaxial scans can be compiled to yield transverse plane, \textit{en face}, images as shown for birefringence (D), OpAx (E) and tractography (F).}
    \label{fig:results:radialEnface}
\end{figure}

By compiling sagittal plane images (uniaxial imaging direction) acquired across the tissue, angular OpAx, birefringence and DOP cross-sections are constructed and are presented in figure~\ref{fig:results:radialEnface}A, \ref{fig:results:radialEnface}B and \ref{fig:results:radialEnface}C, taken from the region indicated by the dashed circle in the transverse (\textit{en face}) view presented in figure~\ref{fig:results:radialEnface}D. These angular views through the cartilage thickness correspond to a circular region with a radius of approximately 2~mm from the centre of the sample, providing a complete 360° view of the loaded cartilage tissue. A description  of how such images are generated is provided in the supplementary material. An important difference seen in these angular views compared to the sagittal plane images (Fig.~\ref{fig:results:opax}B and Fig.~\ref{fig:results:dnDOP}C) obtained from the axisymmetric centre is the absence of the dead-zone in the former because the angular views do not traverse the central axis of loading, where the dead-zone is observed. Also, in the angular views, the SD streak is visible, but in this case is continuous across the entire indented regions (Fig.~\ref{fig:results:radialEnface}B, solid white arrow). In figure~\ref{fig:results:radialEnface}A, it is noted that the OpAx points away from the axisymmetric centre at all angles both above and below the SD line. 

Finally, through the uniaxial imaging direction, we can visualise the projected collagen orientation and birefringence around the entire centre of indentation. A transverse (\textit{en face}) scan, constructed by compiling the sagittal plane images from the uniaxially scanned direction, is shown in figure~\ref{fig:results:radialEnface}D and \ref{fig:results:radialEnface}E. In the indented region, the transverse plane images distinctly shows a central region of low birefringence as also illustrated in the OpAx map (Fig.~\ref{fig:results:radialEnface}D and \ref{fig:results:radialEnface}E). The tractography of this transverse plane image further illustrates the region of low birefringence and also showcases how the OpAx points away from the axisymmetric centre throughout the entire indented region (Fig.~\ref{fig:results:radialEnface}F).  

\clearpage
\section{Discussion}

Firstly, this study demonstrates that both uniaxial and lateral depth-resolved PS-OCT scanning effectively capture structural features that strongly correlate with the microanatomical response of compressed cartilage. Secondly, the combined analysis of OpAx, birefringence, and DOP data provides a means of quantifying the microstructural response of cartilage to loading. Finally, a key finding is that depth-resolved PS-OCT offers a significant advantage over earlier use of cumulative PS-OCT~\cite{goodwin2018quantifying,goodwin2021impact,goodwin2022detection}, to pick out important structural realities in the cartilage response to loading which would otherwise be obscured. This is especially evident in uniaxial imaging, where depth-resolved scans reveal structures that are not present in cumulative PS-OCT images~\cite{goodwin2018quantifying,goodwin2019observing}.

One of the most distinctive structural features observed in this study is the shear discontinuity, a phenomenon extensively described in prior microanatomical studies~\cite{thambyah2006micro,thambyah2007degeneration}. The SD results from lateral shearing within the mid-to-deep zone collagen network. In laterally scanned PS-OCT images, the SD is clearly marked by a shift in optical axis orientation (Fig.~\ref{fig:results:opax}A, white line and Fig.~\ref{fig:results:plots}D), corresponding to a change in the collagen shearing angle above and below the SD. This makes the OpAx image highly informative, as it not only delineates the SD boundary but also provides insight into the surrounding collagen fibril orientation. The potential future work here is to apply different loadings to elicit different shearing responses and to see if there is a consistency in the change in OpAx values.

Another key structural feature is the dead-zone, a region where collagen network shearing bifurcates on either side of the axisymmetric loading centre. In laterally scanned PS-OCT images, the dead-zone is well defined, appearing as a distinct birefringence reduction (Fig.~\ref{fig:results:opax}A, Fig.~\ref{fig:results:dnDOP}A and Fig.~\ref{fig:results:plots}J). This reduction is likely due to a decrease in packed density of the collagen fibrils within the dead-zone. Compared to birefringence and OpAx, the laterally scanned DOP image provides only limited structural information. While it effectively delineates the SD (Fig.~\ref{fig:results:dnDOP}B and Fig~\ref{fig:results:plots}B), it does not capture the dead-zone (Fig.~\ref{fig:results:plots}I). Notably, the DOP image accurately depicts the expected intensification of the SD near the edge of the loaded tissue, indicated by the largest drop in DOP within the SD (Fig.~\ref{fig:results:dnDOP}B, white arrow), a feature not apparent in laterally scanned birefringence or OpAx images.

In the uniaxial view, the OpAx image (Fig.~\ref{fig:results:opax}B) continues to provide a clear representation of the SD and effectively delineates the dead-zone. The presence of fibrillar shearing around the SD is less apparent due to projection effects, which are discussed later. The uniaxial birefringence image (Fig.~\ref{fig:results:dnDOP}C) similarly highlights both the SD and the dead-zone while also illustrating the gradual reduction in fibrillar density within the dead-zone (Fig.~\ref{fig:results:plots}M). However, in contrast to the OpAx and birefringence data, the DOP image (Fig.~\ref{fig:results:dnDOP}D) is the least informative in this view, lacking well-defined structural features (Fig~\ref{fig:results:plots}E and \ref{fig:results:plots}L). Although slight variations in DOP are observed (Fig.~\ref{fig:results:dnDOP}D, white arrow), these changes likely arise from uneven loading of the cartilage tissue rather than distinct microstructural features.

Across both lateral and uniaxial views, the OpAx image offers the best contrast and richest structural detail, making it the most informative modality for assessing cartilage response to compression. While birefringence more effectively characterises the collagen network structure, particularly when considering the SD and dead-zone, the OpAx data instead offers a surrogate measure of the fibril shearing angle, which is ultimately the most useful metric. 

\begin{figure}[tb!]
    \centering
    \includegraphics[width=1\linewidth]{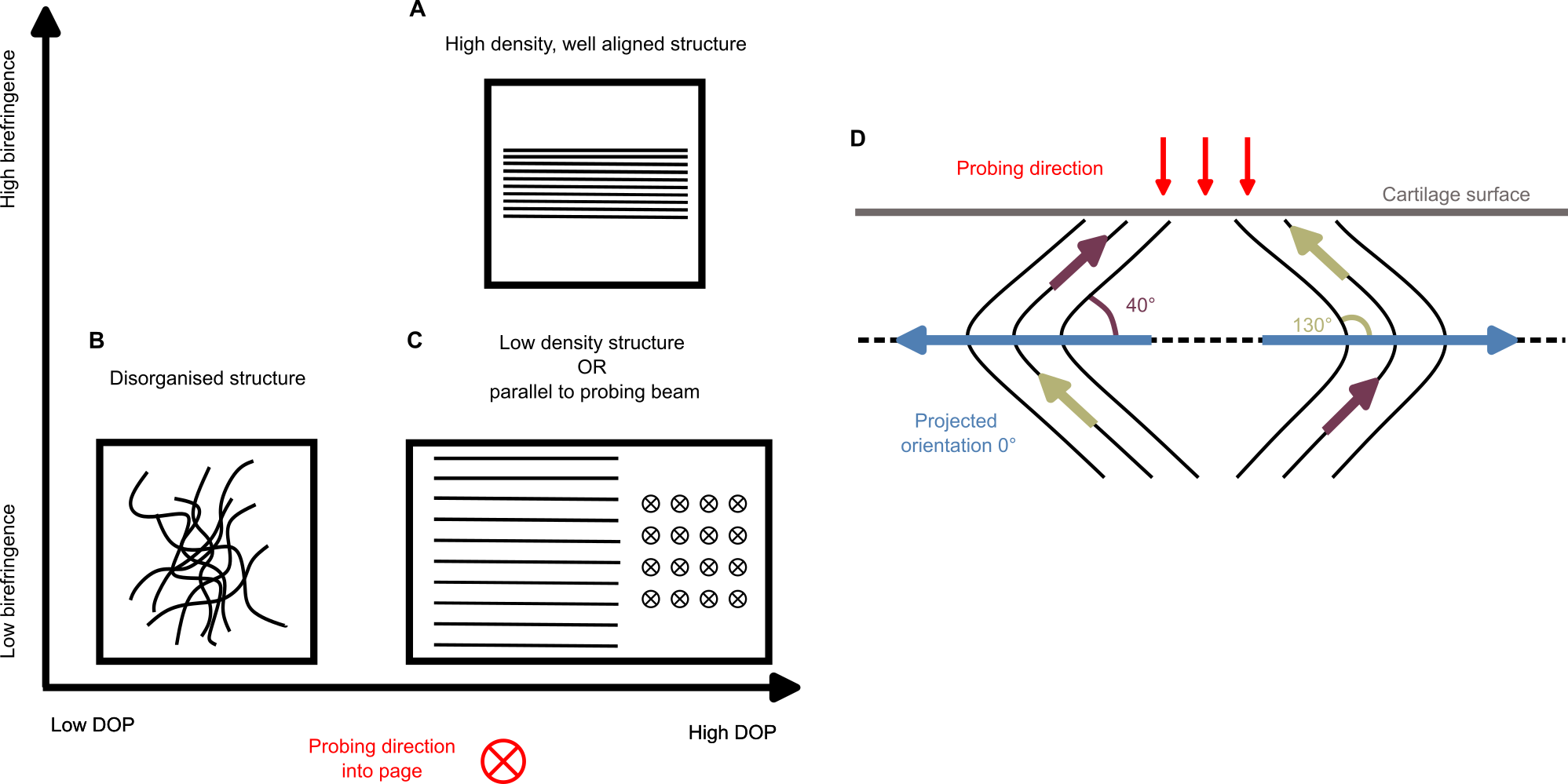}
    \caption{(A-C) Simplified schematic showcasing some of the possible fibrillar architectures depending on level of birefringence and DOP. The laser probing direction is into the page. (D) Diagram showcasing the projection problem in PS-OCT cartilage optical axis measurement. The black lines represent the collagen fibrils forming a `chevron’ pattern in compressed cartilage. The laser probes the cartilage surface to generate a uniaxial scan.}
    \label{fig:discussion:projections}
\end{figure}

Comparatively, the DOP signal is the least informative overall but becomes particularly useful in the dead-zone, where it can help differentiate between fibrillar disorganisation (Fig.~\ref{fig:discussion:projections}B) and density loss (Fig.~\ref{fig:discussion:projections}C). For example, in cases of collagen network disruption due to injury or disease, uniaxial PS-OCT scanning may yield a combined signal of low DOP and low birefringence (Fig.~\ref{fig:discussion:projections}C).

In summary, each polarimetric parameter, the OpAx, birefringence, and DOP, captures distinct aspects of cartilage structure under compression, collectively offering a comprehensive view of the tissue’s mechanical response. Given that all three are derived from the same dataset, we recommend reconstructing them together whenever possible to achieve the most complete interpretation (Fig.~\ref{fig:discussion:projections}A–C). At a minimum, birefringence alone is sufficient to identify both the SD and dead-zone; however, the axisymmetric shear cannot be inferred from birefringence alone. The DOP image alone is only informative in lateral scanning, where it effectively delineates the SD even more clearly than the birefringence and optical axis images, but in the uniaxial direction, it provides little insight into collagen fibril morphology. Importantly, the DOP plays a crucial role in resolving potential degeneracies in interpretation. For instance, low birefringence can arise from multiple structural conditions of the collagen network (Fig.~\ref{fig:discussion:projections}B and \ref{fig:discussion:projections}C), and the DOP may help distinguish between these possibilities. 

Due to the vectorial nature of birefringence, the measured cartilage birefringence and collagen orientation are projected onto a plane orthogonal to the imaging direction (Fig.~\ref{fig:discussion:projections}D). Given the transversely isotropic nature of cartilage, uniaxial imaging provides less detailed structural information compared to lateral imaging. In lateral OpAx images, collagen fibril orientation is well-defined and closely matches DIC microscopy observations.However, the uniaxial view obscures this clarity due to projection effects. For example, in a chevron-like fibril configuration, fibrils oriented at 130° and 40° are both projected onto 0°, resulting in a loss of distinct orientation information (Fig.~\ref{fig:discussion:projections}D, maroon and gold arrows, at 40° and 130° orientations, respectively, both project to the blue 0° arrows). While the exact shearing angle is obscured by projection, the presence of shearing can still be inferred by assessing whether the OpAx values consistently and symmetrically point away from the axisymmetric centre, as observed in the angular image (Fig.~\ref{fig:results:radialEnface}A) and transverse plane image (Fig.~\ref{fig:results:radialEnface}E). For birefringence, the differences between lateral (Fig.~\ref{fig:results:dnDOP}A) and uniaxial (Fig.~\ref{fig:results:dnDOP}C) scans are relatively minor, with most key features remaining visible. However, a notable distinction lies in the dead-zone, where birefringence declines more gradually in lateral imaging compared to the uniaxial view (Fig.~\ref{fig:results:plots}J and \ref{fig:results:plots}M, red oval). Projection effects also influence the DOP scans’ ability to capture the SD in the uniaxial scan direction. However, this loss is minor, as the other polarimetry contrasts effectively depict the SD. 

These projection-related issues highlight the need for a dual-angle PS-OCT~\cite{yao2020high,liu2025three} imaging approach. Recent studies  have demonstrated that probing brain tissue from two different angles allows for the extraction of the full 3D optical axis and absolute sample birefringence~\cite{liu2025three}, effectively eliminating projection artifacts. A logical next step would be to apply a similar method to cartilage, acquiring uniaxial PS-OCT scans at two angles to obtain unprojected polarimetry data, producing optical axis maps comparable to those obtained from lateral imaging but from the uniaxial scan direction.

\begin{figure}[tb!]
    \centering
    \includegraphics[width=1\linewidth]{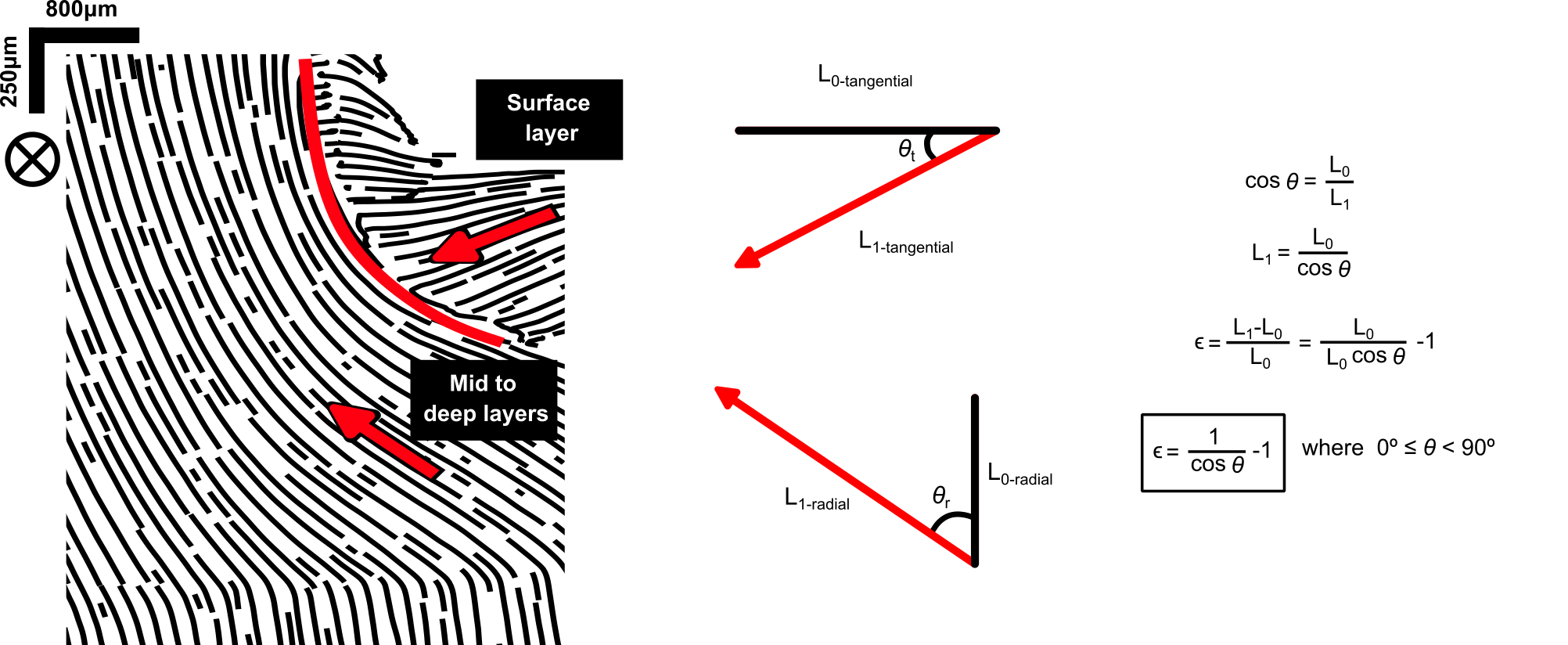}
    \caption{A demonstration of how the tractography information may be used to determine strains across the deformed tissue. The red vectors represent approximate directions of the tractography lines of action. Note the vector deviation from the tangential, $\theta_{\mbox{t}}$, and radial, $\theta_{\mbox{r}}$, orientations in the surface and mid-to-deep zones respectively. This deviation (angle) may then be used to calculate strain, \(\epsilon\).}
    \label{fig:discussion:tractostrain}
\end{figure}

While lateral PS-OCT imaging may initially appear to be merely a lower-resolution alternative to conventional optical microscopy, it offers a particular advantage. Lateral scanning facilitates the generation of accurate, quantitative maps of collagen fibrillar orientation, which in turn supports the application of tractography techniques to visualise and analyse fibrillar morphology (Fig.~\ref{fig:results:tracto}). The unloaded collagen network architecture of adult healthy bovine patellae is well known and has been described in detail previously~\cite{benninghoff1925form,thambyah2006micro}. Typically, the network is tangential in the surface layer or first 10$\%$ thickness of the tissue, and radially aligned in the rest. The junction between the tangential and radial zones is a `transition’ and is likely to be the region that has facilitated the shear boundary between the two zones. By combining the known collagen network architecture in unloaded cartilage, with that of the lines of action in the tractography (Fig~\ref{fig:results:tracto}), we may be able to develop some methods to analyse strain distribution in the tissue. The simple trigonometric relationship between the lines of action relative to the known unloaded state of tangential or radial lines, can potentially be effective to determine relative strain across the tissue (Fig.~\ref{fig:discussion:tractostrain}). 

The structure-based rationale for strain calculation allows for a range of tissue states to be tested with the following assumptions. Firstly, the collagen network is interconnected at the fibrillar level and hence the shear discontinuity is a result of the tangential layer resisting the lateral bulge in the deeper zones. Secondly, it is assumed that the shear pattern is a result of fluid displacement laterally and away from the directly loaded region. Thirdly, the applied stress can be taken as the hydrostatic equilibrium stress, allowing for a stiffness indicator to be obtained by calculating the ratio of stress to strain. 

A particular advantage of uniaxial imaging is its ability to capture the entire loaded region \textit{en face} (Fig.~\ref{fig:results:radialEnface}D–F), allowing for the three-dimensional reconstruction of both the SD and the dead-zone. DIC or other conventional microscopy techniques are unable to easily measure fibrillar deformation across the entire loaded region, highlighting a significant benefit of uniaxial PS-OCT imaging. While lateral scanning can also provide 3D reconstruction, the depth penetration of our PS-OCT system (\textasciitilde2~mm) is considerably smaller than the diameter of the indented region (\textasciitilde5~mm), meaning the entire loaded region is not observed.

The primary advantage of uniaxial scanning PS-OCT over lateral scanning PS-OCT and DIC microscopy is that uniaxial scanning is a completely non-destructive method for assessing cartilage structure under load. While this study used sectioned cartilage, uniaxial PS-OCT can be applied to intact tissue, offering the first method, to our knowledge, for non-destructive observation of both the shear discontinuity and axisymmetric shearing response of cartilage under load. Additionally, we propose that integrating uniaxial depth-resolved PS-OCT with a compression device could allow for dynamic \textit{in situ} imaging of the depth-dependent mechanical response of intact cartilage during loading, without any tissue fixation, providing a physiologically relevant assessment of cartilage function. 

As highlighted in the introduction, cartilage mechanics research has undergone a multi-decade long struggle with two fundamental challenges: (1) the inability to observe the live, dynamic response of cartilage under load and (2) the reliance on destructive techniques for morphological analysis. Depth-resolved PS-OCT uniquely addresses both these long-standing challenges.

\begin{figure}[htb!]
    \centering
    \includegraphics[width=1\linewidth]{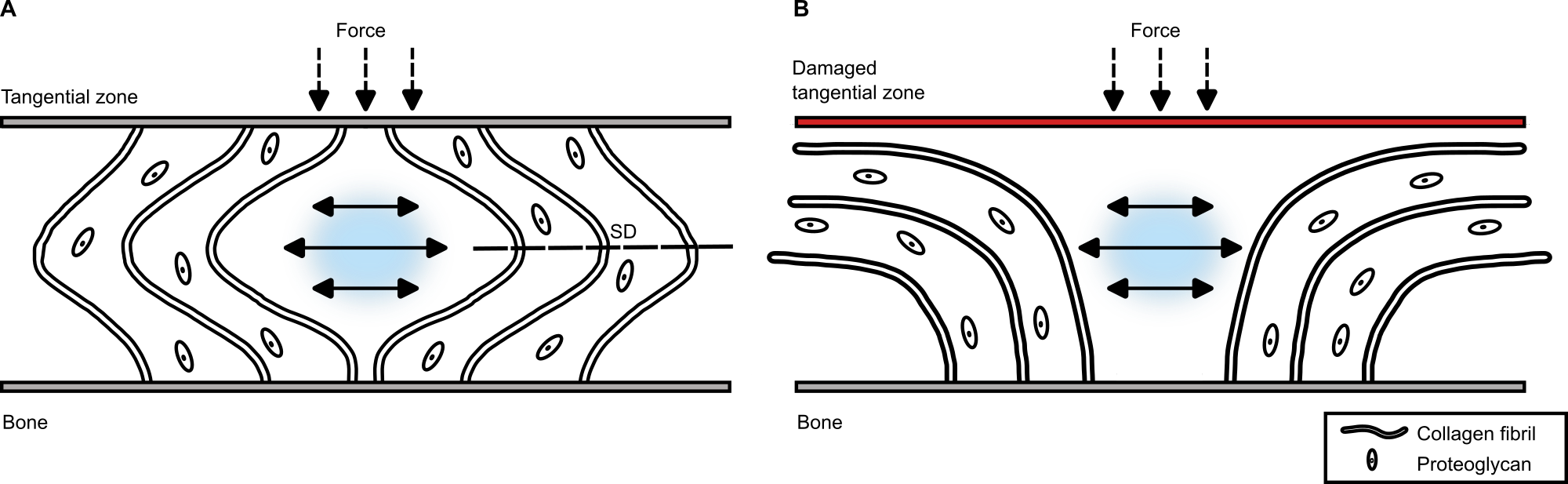}
    \caption{(A) Schematic showcasing the fluid driven formation of the lateral bulge in healthy cartilage. The flow direction is indicated by the black arrows and the dashed arrows indicate the direction of applied force. Shear discontinuity indicated by the dashed black line. (B) Schematic showing the collagen fibrils deformation response when the cartilage tangential layer has been damaged, possibly due to osteoarthritis.}
    \label{fig:discussion:damageDiagram}
\end{figure}

In healthy cartilage, loading induces a well-defined shear discontinuity (Fig.~\ref{fig:discussion:damageDiagram}A), but if the surface layer or mid-to-deep zone is weakened, due to osteoarthritis or structural degradation, the SD may diminish or disappear. This structural change has been observed microstructurally~\cite{thambyah2007degeneration} and would manifest as a `flattening’ of the fibrillar lines of action (Fig.~\ref{fig:discussion:damageDiagram}B). We hypothesize that this response can be observed non-destructively by depth-resolved PS-OCT and used as a marker for early cartilage degeneration, paving the way for future clinical applications.

\section{Conclusion}

The present study has shown that non-destructive imaging of the microstructural response of cartilage to mechanical loading can be achieved using depth-resolved PS-OCT.  This is achieved by combining OpAx, birefringence, and DOP data, to obtain quantifiable indicators of signature cartilage microstructural responses. The present study also determined the extent to which uniaxial scanning is viable compared to lateral scanning, as with the former, there is the most potential for the development of practical instrumentation to assess cartilage health non-destructively \emph{in vivo}. 

\newpage

\section*{Acknowledgements}

The authors acknowledge Dr. Michael Hackmann, Dr. Quingyun Li and Dr. Boy Braaf for insightful discussions that aided in the implementation of the PS-OCT data processing used in this article. Also, we acknowledge funding from the Royal Society Te Apārangi through the Marsden Fund (UOA2110) which made this research possible. We also acknowledge the University of Auckland Doctoral Scholarship, which provided support for Darven Murali Tharan.

\section*{Disclosures}

The authors declare that there are no conflicts of interest related to this article.

\section*{Code, Data, and Materials Availability}

Code and data are available from the corresponding author upon reasonable request. 

\bibliography{Biblio}

\end{document}